\documentclass[aps,prl,reprint,superscriptaddress]{revtex4-1}

\usepackage{amsmath}
\usepackage{graphicx}
\usepackage{upgreek}
\usepackage{color}

\newcommand{\micro}{${\upmu}$}
\newcommand{\per}[1]{{#1}$^{-1}$}

\newcommand{\upsub}[1]{_{\mathrm{#1}}}
\newcommand{\um}{$\,$\micro m}
\newcommand{\uev}{$\,$\micro eV}
\newcommand{\umps}{\um\per{$\,$ps}}
\newcommand{\nlu}{\uev\um$^{2}$}

\begin{document}

\title{Dark Solitons in Waveguide Polariton Fluids Shed Light on Interaction Constants}

\author{P. M. Walker}
\email{p.m.walker@sheffield.ac.uk}
\author{L. Tinkler}
\author{B. Royall}
\affiliation{Department of Physics and Astronomy, University of Sheffield, S3 7RH Sheffield, UK}

\author{D. V. Skryabin}
\affiliation{Department of Physics, University of Bath, BA2 7AY Bath, UK}
\affiliation{ITMO University, Kronverksky Avenue 49, St. Petersburg 197101, Russia}

\author{I. Farrer}
\affiliation{Department of Electronic and Electrical Engineering, University of Sheffield, S3 7HQ Sheffield, UK}

\author{D. A. Ritchie}
\affiliation{Cavendish Laboratory, University of Cambridge, CB3 0HE Cambridge, UK}

\author{M. S. Skolnick}
\author{D. N. Krizhanovskii}

\affiliation{Department of Physics and Astronomy, University of Sheffield, S3 7RH Sheffield, UK}

\begin{abstract}
We study exciton-polariton nonlinear optical fluids in a high momentum regime for the first time. Defects in the fluid develop into dark solitons whose healing length decreases with increasing density. We deduce interaction constants for continuous wave polaritons an order of magnitude larger than with picosecond pulses. Time dependent measurements show a 100ps time for the buildup of the interaction strength suggesting a self-generated excitonic reservoir as the source of the extra nonlinearity. The experimental results agree well with a model of coupled photons, excitons and the reservoir.
\end{abstract}
\maketitle 
Exciton-polaritons are half-light half-matter quasi-particles resulting from strong coupling between photons and quantum-well (QW) excitons~\cite{polaritons}. They behave like photons but experience nonlinearity at least 1000 times larger than in bulk semiconductors due to exciton-exciton scattering~\cite{walker_ncomms}. In a waveguide geometry~\cite{walker_waveguides} the propagation of light is dominated by the high momentum $\beta$ in the propagation direction $z$. The envelope of the optical field evolves slowly compared to the wavelength $2\pi/\beta$ which leads to its evolution equation becoming formally analagous to the Nonlinear Schrodinger (NSE) or Gross Pitaevskii (GPE) equations, but with $z$ playing the role of time~\cite{Agrawal}. This high-momentum paraxial regime has been exploited for photonic simulation of complex Hamiltonians~\cite{Rechtsman_Floquet,Rechtsman_Landau,Suchkov,Longhi,Larre}. Among the most fundamental solutions of the GPE are dark solitons~\cite{
kivshar_solitons,kivshar_dark,Allan,Scwartzlander,Shandarov2000,Amo2011,Grosso,kivshar_finite_extent,proukakis,Frantzeskakis}. These are self-localised dark notches on an infinitely extended bright background accompanied by a phase jump at the center. In the field of nonlinear optics they offer potential applications in all-optical signal processing~\cite{Blair}. The giant polariton nonlinearity allows dark soliton formation at the sub-millimeter length scales needed for on-chip integration~\cite{Amo2011}, a regime previously inaccessible due to weak photon-only nonlinearities. There is, however, still a great deal of experimental uncertainty over the precise nature and strength of the polariton nonlinearity with estimates of the interaction strength varying over two order of magnitude~\cite{Brichkin,ferrier,Rodriguez,Glazov,Vladimirova,Sekretenko}. A proper understanding of these interactions is important as they underpin efforts towards realisation of polariton fermionisation~\cite{Carusotto,Verger,Carusotto_Fermionize} and strongly quantum-correlated states in polaritonic lattices~\cite{Umucalilar,Hafezi_FQH,XChen}.

In this work we experimentally study spatial dark polariton soliton formation in the high momentum regime for the first time. Solitons are formed within 600\um{} at CW powers less than 30mW. We resonantly inject two different classes of initial condition into the waveguides and use the variation of the core size with polariton density to investigate the polariton nonlinearity in the CW regime. We deduce an interaction constant more than an order of magnitude larger than previously observed in the picosecond pulsed regime~\cite{walker_ncomms}. Using time dependent measurements of Gaussian beam self-defocussing we observe that the nonlinearity accumulates on a timescale of order 100ps, much longer than the picosecond response of direct polariton-polariton scattering and consistent with the slow build-up of an excitonic reservoir. Since the polaritons are spatially separated from the pump the reservoir must be generated by the polaritons themselves. Using a numerical model of coherently coupled waveguide photons and excitons and an incoherent reservoir generated by scattering of the excitons we are able to self-consistently fit all experimental features using a single value of the interaction strength. Compared to previous studies conducted using Bragg microcavities~\cite{Amo2011,Grosso}, waveguide propagation always provides a good approximation of the time evolution of the GPE, which enables us to seed dark solitons from different initial conditions. Furthermore, the high momentum allows us to study the density dependent soliton width without the system undergoing the superfluid transition~\cite{Amo2011} and ensures the polariton field is spatially separated from the pump so that we observe only the polariton self-interaction. Finally, the waveguide geometry allows an accurate determination of the number of polaritons injected into the system~\cite{suppl_S2} which allows us to deduce the effective interaction constants. The shrinking of the core size with background density provides an important proof of polariton dark soliton formation, which was previously the subject of some controversy~\cite{cilibrizzi,controv2,controv3}.

\begin{figure}
\centering
\includegraphics[width=8.5 cm]{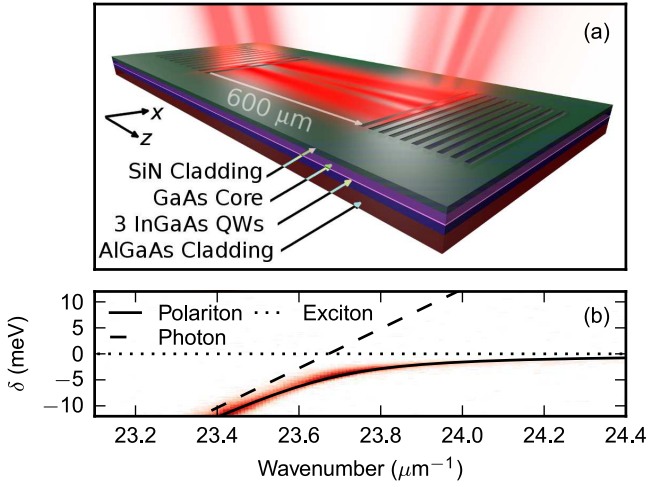}
\caption{(a) Schematic of Polariton Waveguide, (b) Lower polariton dispersion relation seen in angle resolved photoluminescence spectrum.}
\label{fig:schematic}
\end{figure}
The sample used in this work is similar to that in Ref.~\onlinecite{walker_ncomms}. A schematic of the experiment is shown in Figure~\ref{fig:schematic}(a). The polariton dispersion relation is shown in Fig.~\ref{fig:schematic}(b) where the avoided crossing of the uncoupled photon and exciton modes, resulting in a Rabi splitting of 9meV, may be seen. Experiments were performed at 10 Kelvin. A CW laser beam was modified using amplitude or phase masks and then projected onto an input grating coupler (see Fig.~\ref{fig:schematic}(a))~\cite{walker_ncomms,suppl_S1}. The input transverse profile was a 29\um{} FWHM Gaussian with either a phase jump or intensity dip near the center, corresponding to the two classes of initial conditons which we investigate~\cite{suppl_S5}. The polariton fluid undergoes nonlinear evolution in a 600\um{} unpatterned region of planar waveguide and the light was collected by a second grating coupler and imaged onto a CCD camera.
\begin{figure}
\centering
\includegraphics[width=8.5 cm]{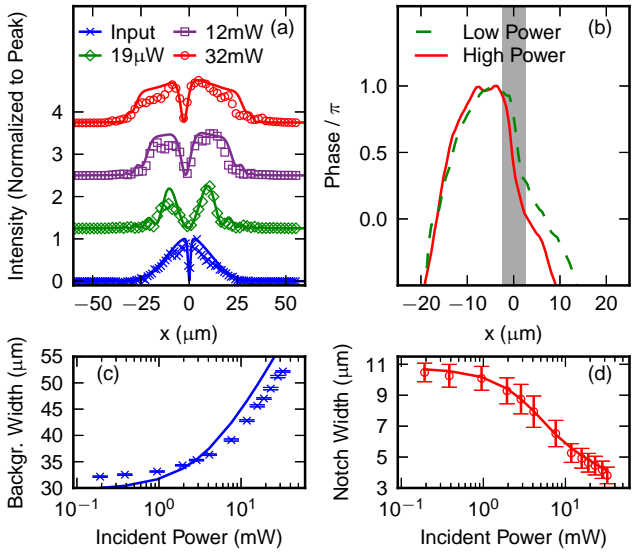}
\caption{Dark soliton formation from phase jump initial condition. (a) Experimental (points) and theoretical (full lines) intensity profiles of input field and output fields at three powers. (b) Experimental output phase at low and high power. (c,d) Experimental (points) and theoretical (full lines) power dependences of full width at one third maximum of background and FWHM of dark notches.}
\label{fig:single}
\end{figure}
We first consider the case of the phase jump initial condition. Figure ~\ref{fig:single}(a) shows the intensity profile of the incident beam and the beam after propagation through the waveguide in the linear and nonlinear regimes. The input profile is Gaussian with a narrow dark notch near the center. At low excitation powers the nonlinearity is negligible and the diffraction of the discontinuous phase in the input field results in a broad dip in the center of the Gaussian background. The background itself is wide enough that it does not experience significant diffraction over the 600\um{} propagation length. Above $\sim$2mW incident power the background broadens under the influence of nonlinear self-defocussing~\cite{Agrawal} while the notch width decreases, eventually forming a single narrow notch in a broad bright background, which is the expected profile for a single dark soliton~\cite{kivshar_solitons}. Figure ~\ref{fig:single}(b) shows a typical phase profile of the output field measured in a separate experiment for low and high power. For both powers a phase jump near $x=0$ is superimposed on a slowly varying background phase arising from the Gaussian background. The shaded region indicates the FWHM of the intensity notch at high power. The phase jump at high power is close to the value of $\pi$ injected at the input~\cite{suppl_S5} as expected in the case of a single dark soliton~\cite{kivshar_solitons}. Figures (c) and (d) show the pump power dependences of the widths of the background and of the dark notch. Crucially, the width of the notch narrows significantly as the density increases, again as expected for a dark soliton. The qualitative reason for this narrowing is the same as for quantised vortex cores in a microcavity polariton condensate~\cite{vortices}. The kinetic energy associated with the localized defect is balanced by the nonlinear potential energy, proportional to the density $n$ of the background. Thus the defect healing length $\xi$ decreases with increasing density according to Eqn.~\eqref{eq:healing_dep}.
\begin{equation}\label{eq:healing_dep}
\xi^2 = v\upsub{g,LP} / \left(2\beta\cos{^2}\phi\cdot n g\right)
\end{equation}
Here $v\upsub{g,LP}$=24\umps{} and $\beta$=23.7\per{\um{}} are the polariton group velocity and wavenumber, $\phi$ is the soliton phase angle and $\hbar g$ is the polariton-polariton interaction energy per unit polariton density. Note that $\hbar\beta/v\upsub{g,LP}$ plays the role of mass and that the core FWHM$\approx$2.493$\xi$ \cite{suppl_S4}. Eqn.~\eqref{eq:healing_dep} follows from the analytical dark soliton solution of the GPE~\cite{suppl_S4,kivshar_solitons,kivshar_dark}.
\begin{figure}
\centering
\includegraphics[width=8.5 cm]{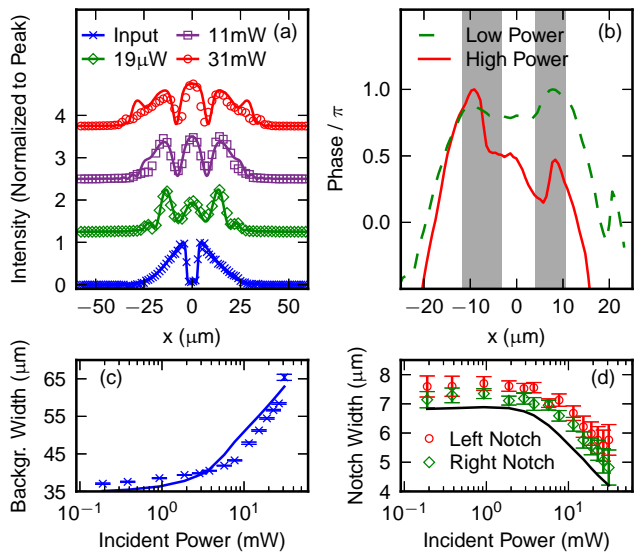}
\caption{As Fig.~\ref{fig:single} for amplitude defect initial condition.}
\label{fig:double}
\end{figure}
In the case of the amplitude defect initial condition the field profiles are shown in figure ~\ref{fig:double}(a). The intensity notch is clearly seen in the input field. At high density a pair of well defined dark notches is present on the broad bright background of the output field. Figure ~\ref{fig:double}(b) shows the difference in output phases between low and high power cases. An important evidence of dark soliton formation is that there is a modification of the phase profile at high densities such that two phase jumps of opposite sign occur at the positions of the dark notches (shaded grey) which are not present at low power. This shows that the nonlinear interaction has generated the correct phase profile for a pair of dark solitons seeded by an intensity defect in the initial condition~\cite{kivshar_solitons}. Furthermore, as shown in Fig.~\ref{fig:double}(d), the notch widths again decrease with increasing density as expected for a solitonic solution.
\begin{figure}
\centering
\includegraphics[width=8.5 cm]{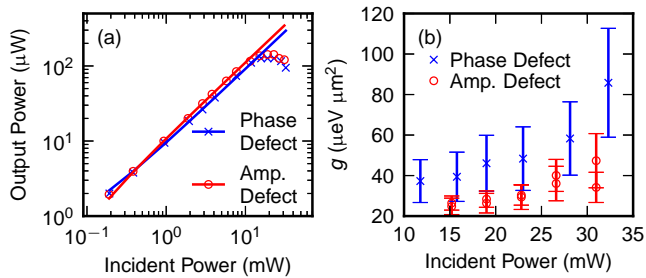}
\caption{(a) Output vs. Incident power for the two initial conditions. (b) Nonlinearity deduced from healing length.}
\label{fig:power}
\end{figure}
We now consider the origin and size of the nonlinear interaction responsible for generating the solitons. Figure ~\ref{fig:power}(a) shows the measured output power from the waveguide vs. the incident power, which is linear up to 10mW where the narrowing of the dark notches and broadening of the background is already pronounced. From the gradient and known losses~\cite{walker_ncomms} we determine that 20$\pm$2\% of the incident light couples into the guided mode~\cite{suppl_S2}. A first estimate of the effective size of the polariton-polariton interaction constant $g$ may be made from Eqn.~\eqref{eq:healing_dep} using the peak polariton density $n$. We use $\cos\phi=$1 for the phase defect and 0.9 for the amplitude defects, deduced from the valley-to-peak ratio of the dark notches~\cite{suppl_S4}. Values of $g$ for several powers are shown in Fig.~\ref{fig:power}(b) and are in the range 25-37\nlu{} for powers where the output vs. input power is linear. The amplitude and phase defects give very similar values. We note that for a dark soliton with core FWHM 7\um{} the balanced nonlinear and diffraction lengths are equal to $2\beta\xi^{2}$=370\um{}, which is less than the device length $L$=600\um{}. We can therefore be sure that solitons narrower than this are able to form in our device~\cite{suppl_S4}. The values of $g$ we obtain are more than two orders of magnitude larger than $g$=0.3\nlu{}, which we previously deduced for dark and bright solitons using very similar structures but with picosecond pulses~\cite{walker_ncomms}. Scaling to account for the different detuning $\delta$=-7.6meV in those measurements gives $g$=1.5\nlu{}, still over an order of magnitude less than we see here. This may occur if some of the polaritons are transferred into an excitonic reservoir with a lifetime long compared to the picosecond pulses. In the steady state the reservoir population will be larger than that of strongly-coupled excitons by the ratio of its lifetime to excitation time, providing an excess exciton population which makes the polariton interaction appear stronger than it really is. The timescale of picosecond pulses is short compared to the reservoir excitation rate so no significant extra population builds up and only the strongly coupled exciton population contributes to the nonlinearity. The reservoir may be generated by scattering from disorder into excitonic states in the tail of the inhomeogeneously broadened exciton line~\cite{Whittaker}, for which the density of states is much larger than for polaritons. Alternatively, the reservoir can be composed of localised, indirect or otherwise dark excitons~\cite{Krizh_localised,Menard_dark_excitons,Sarker_dark_excitons,Cundiff,Snoke_spin_flip}.
\begin{figure}
\centering
\includegraphics[width=8.5 cm]{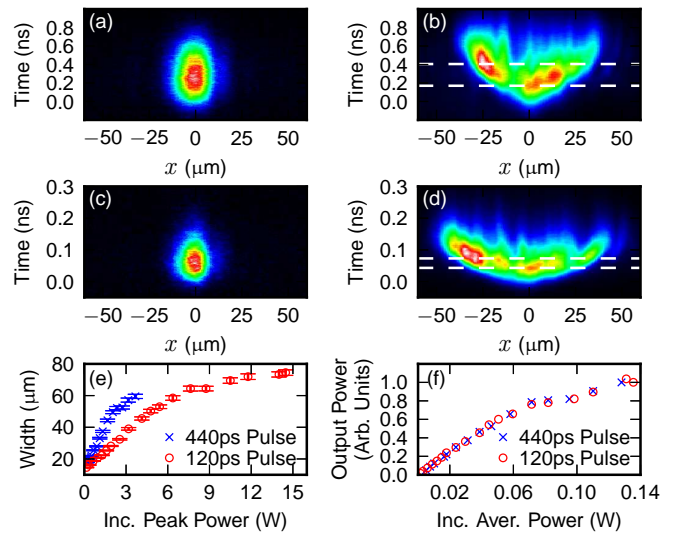}
\caption{Spatial defocussing of long pulses. Space-time plots of output intensity for incident pulses with temporal FWHM and peak powers (a) 440ps, 0.14W (b) 440ps, 3.7W (c) 120ps, 0.16W (d) 120ps, 14W. (e) Width of time-integrated spatial distribution vs. incident peak power. (f) Integrated output intensity vs. input power.}
\label{fig:long_pulses}
\end{figure}

To further investigate this effect we injected pulses of length 120ps and 440ps with a gaussian spatial profile into the waveguide and observed the spatial defocussing as a function of time using a streak camera. The pulses were detuned $\hbar\delta =$-7.2meV from the exciton and the spatial FWHM was 15\um{}. Fig.~\ref{fig:long_pulses} shows the output intensity as a function of $x$ and time $t$. In the low power case (a,c) the output pulse is unchanged and is gaussian in both $x$ and $t$. At high power (b,d) the spatial distributions broaden as previously observed in the CW case. It can be seen that the outer portions of the spatial distribution arrive delayed with respect to the center. Taking sections at $\pm$25\um{} the delay is 30ps in the case of 120ps long pulses and 240ps in the case of 440ps pulses. These delays are too large to be explained in terms of a change in velocity as the travel times for polaritons at this detuning and for pure photons are 14ps and 10ps respectively. In Fig.~\ref{fig:long_pulses}(e) it can be seen that the spatial width of the distribution increases twice as fast with peak pulse power in the case in of 440ps pulses compared to 120ps pulses. The coupling efficiency, absorption, etc. are the same for both pulse lengths as seen from the almost identical input vs. output power curves in Fig.~\ref{fig:long_pulses}(f). This implies that the longer pulses experience twice the nonlinearity. Taken together, these two effects can be explained if polaritons in the first part of the pulse generate a reservoir which increases the nonlinear interaction for the latter part of the pulse resulting in increased spatial defocussing at later times.

We now consider a numerical model which provides a self-consistent fit to all features of the experimental data. Polaritons propagating in the waveguide are described by Eqns.~\eqref{eq:fields} for the slowly varying coupled photon and exciton envelope amplitudes $A$ and $\psi$ and the reservoir density $n\upsub{R}$.
\begin{subequations} \label{eq:fields}
\begin{gather}
\left(i\frac{\partial}{\partial t} + i\gamma\upsub{p} + v\upsub{g}\left(i\frac{\partial}{\partial z}+ \frac{1}{2\beta\upsub{e}}\frac{\partial^2}{\partial x^2}\right)\right)A = \left(\frac{\Omega}{2}\right)\psi \\
\left(i\frac{\partial}{\partial t} + i\left(\gamma\upsub{e} + \gamma\upsub{r}\right) - g\upsub{X}\left(\left|\psi\right|^2 + n\upsub{R}\right)\right)\psi = \left(\frac{\Omega}{2}\right)A\label{eq:fields_X}\\
\frac{\partial n\upsub{R}}{\partial t} = 2\gamma\upsub{r}\left|\psi\right|^{2} - 2\gamma\upsub{R}n\upsub{R} \label{eq:fields_res}
\end{gather}
\end{subequations}
Here, $v\upsub{g}$=58\umps{} and $\beta\upsub{e}$=23.6\per{\um{}} are the photon group velocity and wavenumber at the exciton frequency extracted from the fit to the dispersion relation in Fig.~\ref{fig:schematic}(b). The loss rates $\gamma\upsub{p}$ and $\gamma\upsub{r}$ are due to photon tunnelling through the cladding and loss of excitons due to scattering to the reservoir while $\gamma\upsub{e}$ quantifies all other exciton loss channels. The reservoir decay rate is $\gamma\upsub{R}$. The total homogeneous exciton linewidth $\hbar\left(\gamma\upsub{e}+\gamma\upsub{r}\right)$=13.2\uev{} and $\hbar\gamma\upsub{p}$=32.9\uev{} were obtained from an independent fit to the spectral dependence of the loss length~\cite{suppl_fig_S1}. The Rabi splitting $\hbar\Omega$=9meV and the (polarisation averaged) exciton interaction energy per unit exciton density in one QW is given by $\hbar g\upsub{X}$.

In the steady state, where $\partial n\upsub{R}/\partial t$=0, Eqn.~\eqref{eq:fields_res} can be rearranged to give $n\upsub{R}=(\gamma\upsub{r}/\gamma\upsub{R})\left|\psi\right|^{2}$. Substituting this into Eqn.~\eqref{eq:fields_X}, the nonlinearity $g\upsub{X}(\left|\psi\right|^2 + n\upsub{R})$ becomes $g\upsub{eff}\left|\psi\right|^2$ where $g\upsub{eff}=g\upsub{X}\left(1+\gamma\upsub{r}/\gamma\upsub{R}\right)$ is an effective exciton-exciton scattering which accounts for the fact that for every strongly coupled exciton the reservoir contains another $\gamma\upsub{r}/\gamma\upsub{R}$ incoherent excitons. For CW driving we use the ansatz $A = A\left(x,z\right)\exp\left(-i\delta t\right)$, and likewise for $\psi$, and eliminate $\psi$ using Eqn.~\eqref{eq:fields_X} to leave a generalised GPE for $\partial A\left(x,z\right)/\partial z$ in terms of $A$ ~\cite{suppl_S3}. This was solved using a standard split-step Fourier method~\cite{Agrawal}. The input conditions are well determined as evidenced by the good agreement between experiment and theory at low power (Figs.~\ref{fig:single},~\ref{fig:double}(a) and~\cite{suppl_S5}).

The model output is plotted as solid lines in figures~\ref{fig:single},\ref{fig:double}(a,c,d). Good semi-quantitative agreement is achieved with all intensity profiles (panels (a)) and with the dark notch and background widths (panels (c,d)) at all powers. The only adjustable parameter $\hbar g\upsub{eff}$=220\nlu{} fits all the above data at once so the model provides a self-consistent explanation of all features. The effective polariton-polariton interaction $g$ corresponding to the above exciton-exciton interaction is obtained using $g$=$g\upsub{eff}\left|X\right|^4/N\upsub{w}$ where $\left|X\right|^2$=0.58 is the exciton fraction and $N\upsub{w}$=3 is the number of QWs~\cite{Brichkin}. This gives $\hbar g$=25\nlu{}, in good agreement with the value deduced from the soliton healing length. Thus the numerical model is also consistent with the picture of dark solitons with healing length given by Eqn.~\eqref{eq:healing_dep}. If we take $\hbar g\upsub{X}$=13.5\nlu{} from our work with picosecond pulses~\cite{walker_ncomms} then to obtain the observed $g\upsub{eff}$ a ratio of 15 between the reservoir lifetime and excitation rates is required. Since the total exciton loss rate $\gamma\upsub{e}+\gamma\upsub{r}=$13.2\uev{} corresponds to 25ps the reservoir lifetime is longer than 375ps. Considering the time resolved data, the reservoir excitation time of order 25ps is short compared to the pulse lengths so the reservoir effect can also explain the enhanced nonlinearity felt by the trailing edge of the pulses. Comparing our value with others in the literature, Rodriguez et. al. deduced $g\upsub{eff}$=30\nlu{}, which lies between our picosecond and CW reults, in a similar system to ours. Ferrier et. al. \cite{ferrier} quote $g$=2-9\nlu{} for polaritons spatially separated from the pump from which we infer $g\upsub{eff}\sim$50-225\nlu{}. In a similar experiment Brichkin et. al.~\cite{Brichkin} find $g\upsub{eff}$=2.4\nlu{}, of the same order as theoretical estimates. The differences between similar systems suggests a strong dependence of the CW nonlinearity on sample properties. Our picture of a reservoir generated by the polaritons themselves may explain this behaviour since variations in QW disorder between different semiconductor wafers can strongly influence scattering into the reservoir.

In conclusion, we have shown dark soliton formation from both amplitude and phase defects in a polariton fluid and observed the density dependence of the soliton healing length for the first time. We have measured a polariton-polariton interaction strength more than an order of magnitude larger than polariton interactions on a picosecond timescale and shown that the extra nonlinearity develops over a timescale of order 100ps. We attribute it to the slow build-up of a reservoir excited by the polaritons, which contributes to the blueshift of the coherent exciton field. Our numerical model explains all experimental features for both initial conditions using a single value of the excitonic interaction strength. We thus achieve a self-consistent picture of dark soliton formation and CW polariton interactions.
\begin{acknowledgments}
We acknowledge support from EPSRC Programme Grant EP/J007544/1, ERC Advanced Grant 320570 and Leverhulme Trust Grant PRG-2013-339. PMW acknowledges helpful discussions with D. M. Whittaker.  
\end{acknowledgments}
%

\end{document}